\documentclass[prl,twocolumn,showpacs,superscriptaddress]{revtex4}

\usepackage{graphicx}
\usepackage{amsmath}
\usepackage{amsfonts}
\usepackage{bm}
\usepackage{bbm}

\begin{document}


\title{All-optical switching in planar semiconductor microcavities}

\author{S. Schumacher}
\author{N. H. Kwong}
\author{R. Binder}

\affiliation{College of Optical Sciences,
             University of Arizona,
             Tucson, Arizona 85721, USA}

\author{Arthur L. Smirl}

\affiliation{Laboratory for Photonics and Quantum Electronics, 138
IATL, University of Iowa, Iowa City, Iowa 52242, USA}

\date{\today}

\newcommand{\ME}{{\genfrac{}{}{0.0pt}{3}{\text{TM}}{\text{TE}}}}
\newcommand{\EM}{{\genfrac{}{}{0.0pt}{3}{\text{TE}}{\text{TM}}}}
\newcommand{\TM}{\text{TM}}
\newcommand{\TE}{\text{TE}}
\newcommand{\probe}{\text{probe}}
\newcommand{\pump}{\text{pump}}
\newcommand{\probepump}{{\genfrac{}{}{0.0pt}{3}{\text{probe}}{\text{pump}}}}

\newcommand{\XY}{{\genfrac{}{}{0.0pt}{3}{\text{X}}{\text{Y}}}}
\newcommand{\YX}{{\genfrac{}{}{0.0pt}{3}{\text{Y}}{\text{X}}}}
\newcommand{\X}{\text{X}}
\newcommand{\Y}{\text{Y}}

\pacs{190.4380, 320.7110, 190.3100}

\begin{abstract}
Using a microscopic many-particle theory, we propose all-optical
switching in planar semiconductor microcavities where a weak beam
switches a stronger signal. Based on four-wave-mixing instabilities,
the general scheme is a semiconductor adaptation of a recently
demonstrated switch in an atomic vapor [Dawes et al., Science 308,
672 (2005)].
\end{abstract}

\maketitle

\textsc{Introduction} -- In a recent study, an all-optical switch
operating at very low light intensities was demonstrated in an
atomic vapor system \cite{Dawes2005}. For two pump pulses
counter-propagating through an atomic vapor, four-wave mixing (FWM)
induced (spontaneous) off-pump-axis pattern formation can occur
\cite{Yariv1977,Grynberg1988,Chang1992a}. With a slight breaking of
the system symmetry to impose a preferred direction to this off-axis
pattern formation, it was demonstrated that a very weak ``probe''
pulse is sufficient to shift the patterns away from the preferred
direction into the probe direction \cite{Dawes2005}. In this way, an
all-optical switch was realized, where the switched signal is much
stronger than the control pulse (in this case the probe).

A semiconductor version of that switch is desirable, and would be
potentially useful in all-optical communication systems. In this
Letter, we propose a specific implementation of such a switch. At
the heart of our proposal is a planar semiconductor microcavity, in
which FWM induced instabilities are already well established (see
e.g.,
\cite{Savvidis2000,Ciuti2000,Saba2001,Whittaker2001,Diederichs2006,Romanelli2007}).
The physical processes underlying instabilities in microcavities are
quite different from instabilities found in atoms, and therefore, it
is not trivial to transfer the atomic concepts to a semiconductor
switch. In atoms, the dominant optical nonlinearity leading to FWM
and instability is saturation (a process usually referred to as
phase-space filling (PSF) in the semiconductor community). While in
semiconductor microcavities, it is the repulsive Coulomb interaction
between cavity polaritons. Despite the differences with atomic
systems in terms of physics and geometry, we show that the
microcavity still permits the same basic switching concept.

\begin{figure}[t]
{\begin{center}
\includegraphics[scale=1.0]{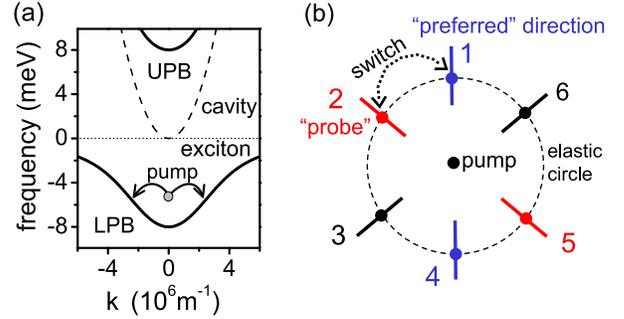}
\end{center}}
\caption{\label{dispersion}(Color online) (a) Sketch of the linear
cavity polariton dispersion. The bare cavity and exciton dispersions
are shown, together with the lower (LPB) and upper (UPB) polariton
branches of the coupled cavity-mode exciton system. The fundamental
pairwise off-axis scattering of pump polaritons is also indicated.
(b) Sketch of the hexagonal switching geometry in the transverse
plane. The elastic circle is defined by the pump frequency and the
dispersion of the LPB. The basic switching action triggered by the
probe is indicated. The radial bars indicate the variation in the
magnitude of off-axis momenta $\mathbf{k}$ as included in the
nonlinear polariton dynamics.}
\end{figure}

\textsc{Theory} -- We analyze the nonlinear cavity-polariton
dynamics in a typical planar GaAs microcavity
\cite{Schumacher2007a}. The linear polariton dispersion is shown in
Fig.~\ref{dispersion}(a). Our theory is a $\chi^{(3)}$ Hartree-Fock
(HF) theory in the coherent limit, evaluated self-consistently up to
arbitrary order in the optical fields
\cite{Buck2004,Schumacher2005a,Oestreich1999,Schumacher2006}. The
polariton interactions are strongly spin-dependent
\cite{Kavokin2003,Krizhanovskii2006,Schumacher2007a,Martin2007}.
Here we concentrate on the dynamics in one spin subsystem (say spin
up) by choosing circularly polarized excitation. We neglect a
possible longitudinal-transverse (TE-TM) cavity-mode splitting
\cite{Savona1996}. We apply a spatial decomposition of cavity field
and exciton polarization into Fourier components $E_{\mathbf{k}}$
and $p_{\mathbf{k}}$, respectively, with in-plane momentum
$\mathbf{k}$ \cite{Savasta2003}. The exciton dynamics is evaluated
in the 1s heavy-hole exciton subspace. The nonlinear set of coupled
equations of motion for  $E_{\mathbf{k}}$ and $p_{\mathbf{k}}$ reads
\cite{Savasta2003,Schumacher2007a}:
\begin{align}
\label{fieldmotion}
i\hbar\dot{E}_{\mathbf{k}}=&\hbar\omega_{\mathbf{k}}^cE_\mathbf{k}
-\Omega_{\mathbf{k}}p_\mathbf{k}+i\hbar
t_cE^{\text{eff}}_{\mathbf{k},\text{inc}}\,, \\
\label{generalmotion}
i\hbar\dot{p}_{\mathbf{k}}=&\big(\varepsilon_{\mathbf{k}}^x-i\gamma_x\big)p_{\mathbf{k}}-\Omega_{\mathbf{k}}E_{\mathbf{k}}
\nonumber
+\sum_{\mathbf{q}\mathbf{k}'\mathbf{k}''}{\big(2\tilde{A}\Omega_{\mathbf{k}''}p_{\mathbf{q}}^\ast}p_{\mathbf{k}'}E_{\mathbf{k}''} \\
&+V_{\text{HF}}{p^\ast_{\mathbf{q}}p_{\mathbf{k}'}p_{\mathbf{k}''}\big)\delta_{\mathbf{q},\mathbf{k}'+\mathbf{k}''-\mathbf{k}}}\,.
\end{align}
The cavity-field in Eq.~(\ref{fieldmotion}) is treated in quasi-mode
approximation. The effective incoming field
$E^{\text{eff}}_{\mathbf{k},\text{inc}}$ driving the field
$E_\mathbf{k}$ in the cavity mode is obtained from a simple
transfer-matrix formalism that includes the radiative width
($\Gamma=\frac{\omega\hbar^2t_c^2}{\varepsilon_0cn_{\text{b}}}$,
with the background refractive index $n_{\text{b}}$, the vacuum
velocity of light $c$ and dielectric constant $\varepsilon_0$) of
the cavity mode and yields transmitted and reflected field
components:
$E^{\text{eff}}_{\mathbf{k},\text{inc}}=E_{\mathbf{k},\text{trans}}-E_{\mathbf{k},\text{refl}}$
with
$E_{\mathbf{k},\text{trans}}=E_{\mathbf{k},\text{inc}}+\frac{\hbar
t_c}{2n_{\text{b}}c\varepsilon_0}\dot{E}_{\mathbf{k}}$ and
$E_{\mathbf{k},\text{refl}}=-\frac{\hbar
t_c}{2n_{\text{b}}c\varepsilon_0}\dot{E}_{\mathbf{k}}$. The
cavity-to-outside coupling constant $t_c$ is chosen such that
$\Gamma\approx0.4\,\mathrm{meV}$ for
$\hbar\omega=1.5\,\mathrm{meV}$. We include excitonic PSF and HF
exciton-exciton Coulomb interaction in the nonlinear exciton
dynamics in Eq.~(\ref{generalmotion}); two-exciton correlations are
neglected in this study and are expected to give merely quantitative
changes because the pump is tuned far (several $\mathrm{meV}$) below
the bare exciton resonance \cite{Kwong2001b,Savasta2003} (cf.
Fig.~\ref{dispersion}(a)). Inclusion of two-exciton Coulomb
correlations in our calculations would basically lead to
renormalization of $V_{\text{HF}}$ in Eq.~(\ref{generalmotion}) and
give rise to a small additional excitation-induced dephasing
\cite{Kwong2001b,Savasta2003}. The bare exciton and cavity in-plane
dispersions are denoted by $\varepsilon^x_{\mathbf{k}}$ (with
$\varepsilon^x_{{0}}=1.497\,\mathrm{eV}$) and
$\omega^c_{\mathbf{k}}$, with
$\hbar\omega^c_{\mathbf{k}}=\varepsilon^x_{{0}}/\cos{\vartheta}$ and
$\sin{\vartheta}=|\mathbf{k}|c/(\omega n_{\text{b}})$. A
phenomenological dephasing constant $\gamma_x=0.4\,\mathrm{meV}$ is
included for the exciton polarization,
$\Omega_{\mathbf{k}}=8\,\mathrm{meV}$ is the vacuum Rabi splitting,
and $\tilde{A}=A_{\text{PSF}}/\phi_{1s}^\ast(0)$ (with
$A_{\text{PSF}}=4a_0^x\sqrt{2\pi}/7$, the two-dimensional exciton
Bohr radius $a_0^x\approx170\,\AA$, and the two-dimensional 1s
exciton wavefunction $\phi_{1s}(\mathbf{r})$ at $\mathbf{r}=0$) and
$V_{\text{HF}}=2\pi(1-315\pi^2/4096)/a_0^{x2}E_b^x$ (with the bulk
exciton binding energy $E_b^x\approx13\,\mathrm{meV}$) are the
excitonic PSF and HF Coulomb matrix elements, respectively. A
spatial anisotropy in the system can be modeled, e.g., by including
an anisotropic cavity dispersion $\omega_{\mathbf{k}}^c$.

\begin{figure}[t]
\begin{center}
\includegraphics[scale=1.0]{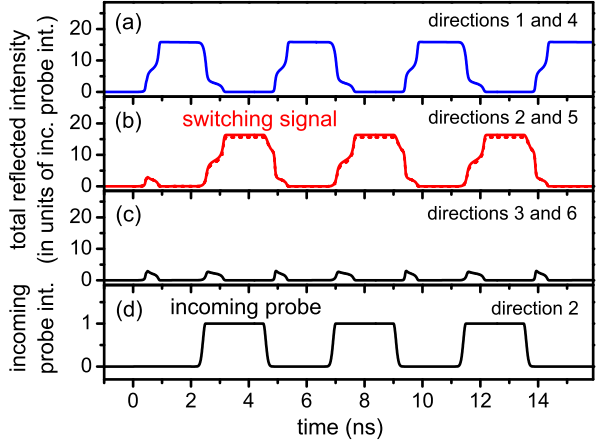}
\caption{\label{Figswitch}(Color online) (a)-(c) Switching in the
reflected signals. The intensities per direction are normalized to
the incoming probe intensity. The switching signal in (b) is about
$15$ times stronger than the incoming probe in (d) that is
triggering this signal (note the different scales on the vertical
axes in panels (a)-(c) and (d)). In panel (b), direction 2 is shown
as the solid line and direction 5 as the dashed line. Similar
switching is observed in transmission (not shown).}
\end{center}
\end{figure}

\begin{figure}[t]
\begin{center}
\includegraphics[scale=1.0]{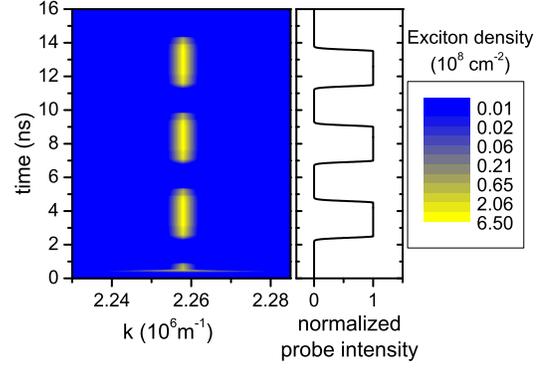}
\end{center}
\caption{\label{Figtimetrace1}(Color online) Timetrace of the
exciton density in direction 2 in which the probe pulse is applied.}
\end{figure}

We consider steady-state pump excitation in normal incidence
($\mathbf{k}_{\text{pump}}=0$), spectrally below the bare exciton
resonance and above the lower polariton branch (cf.
Fig.~\ref{dispersion}(a)). For this excitation configuration, FWM
processes triggered by fluctuations in the cavity photon field can
give rise to resonant and phase-matched (momentum and energy
conserving) pairwise off-axis scattering of pump-excited polaritons
into two polaritons with finite and opposite in-plane momentum
$\mathbf{k}$ and $-\mathbf{k}$ (cf. Fig.~\ref{dispersion}(a)). For a
pump-induced exciton density above the instability threshold
($\gtrsim10^{10}\,\mathrm{cm^{-2}}$ here) these scattering processes
can lead to strong off-axis signals as has been demonstrated in
Ref.~\onlinecite{Romanelli2007} for a slightly different excitation
geometry. With a small symmetry breaking (anisotropy) in the system,
hexagons and their subsets were shown to be the favored
instability-induced patterns in \cite{Dawes2005}.

In Eqs.~(\ref{fieldmotion}) and (\ref{generalmotion}), we allow for
signals in the pump direction ($\mathbf{k}_{\text{pump}}=0$) and six
off-axis directions forming hexagons in the transverse plane (cf.
Fig.~\ref{dispersion}(b)). Contributions to $E_{\mathbf{k}}$ and
$p_{\mathbf{k}}$ with finite in-plane momentum $\mathbf{k}\neq0$
(off-axis contributions) are restricted to
$k_{\text{min}}<|\mathbf{k}|<k_{\text{max}}$. The lower
($k_{\text{min}}\approx2.0\times10^6\,\mathrm{m^{-1}}$) and upper
($k_{\text{max}}\approx2.7\times10^6\,\mathrm{m^{-1}}$) momentum
cut-offs are chosen such that the elastic circle (including the
dynamically changing nonlinear renormalizations) lies within the
numerical domain. This approximation can be justified as long as the
pump excitation takes place spectrally below the ``magic angle''
(analog to the experiments in \cite{Romanelli2007}). Otherwise,
energy and momentum conserving resonant scattering could
significantly populate off-axis states away from the elastic circle.

\textsc{Results \& Discussion} -- In Figs.~\ref{Figswitch} and
\ref{Figtimetrace1}, we show results where we have numerically
integrated the coupled Eqs.~(\ref{fieldmotion}) and
(\ref{generalmotion}) for quasi steady-state pump excitation in
normal incidence. The pump frequency is tuned $5\,\mathrm{meV}$
below the bare exciton resonance. The pump (not shown) reaches its
peak intensity $I_{\text{pump}}\approx19.5\,\mathrm{kWcm^{-2}}$
shortly after $0\,\mathrm{ps}$ and is then kept constant. We impose
a slight anisotropy in the cavity dispersion by shifting
$\omega_{\mathbf{k}}^c$ to lower energies by $0.075\,\mathrm{meV}$
in directions 1 and 4. Above a certain pump threshold intensity,
phase-matched pairwise scattering of pump-induced polaritons, driven
mainly by the HF term in Eq.~(\ref{generalmotion}), leads to
spontaneous (fluctuation-triggered) off-axis signal formation
(similar to \cite{Romanelli2007,Verger2007}). Initially, signals in
all the considered off-axis directions start to grow simultaneously.
However, as these signals grow over time, the anisotropy (symmetry
breaking) fixes the spontaneous off-axis pattern at directions 1 and
4. This can be seen in Fig.~\ref{Figswitch}(a)-(c) for times less
than $2\,\mathrm{ns}$. After $2\,\mathrm{ns}$, we apply a weak probe
($I_{\text{probe}}\approx0.1\,\mathrm{Wcm^{-2}}$) with the same
frequency as the pump frequency in direction 2
(Fig.~\ref{Figswitch}(d)). Now, the strong off-axis emission
switches to directions 2 and 5 and vanishes in the ``preferred''
directions 1 and 4. Note that the switching signal in directions 2
and 5 is about \emph{15 times stronger} than the probe pulse itself
(i.e., part of the pump is redirected from normal incidence to the
directions 2 and 5). In other words, the gain in direction 2 is
$\approx11.7\,\mathrm{dB}$. When switching off the weak probe at
$\approx5\,\mathrm{ns}$, the strong off-axis emission switches back
to the preferred directions 1 and 4. The switching can then be
repeated as shown in the figure. Finally, Fig.~\ref{Figtimetrace1}
shows the time evolution of the exciton density in direction 2
during the switching process. The density is resolved into its
in-plane momentum components. For the chosen parameters the total
off-axis density reaches about $10\%$ of the pump-induced exciton
density.

Figures~\ref{Figswitch} and \ref{Figtimetrace1} demonstrate the
switching for one specific set of parameters. However, the general
mechanism is robust. By fine-tuning the parameters in the relatively
large parameter space that determines the system dynamics (e.g.,
pump and probe intensities and frequencies, and probe in-plane
momentum), the switching performance shown in Figs.~\ref{Figswitch}
and \ref{Figtimetrace1} can be optimized further, e.g., to achieve
larger gain. Since the total strength of the off-axis signals is
mainly determined by the pump excitation parameters, the largest
gain is typically achieved for the lowest probe intensity that can
trigger the switching process. We also note that the system dynamics
drastically changes when the pump intensity and/or frequency is
chosen such that bistability \cite{Gippius2004,Wouters2007} plays a
role.

Finally, we briefly discuss possible limitations of the proposed
scheme. Since the pump excitation is off-resonant, a relatively
strong pump is required to reach the instability threshold. In an
experimental setup, unintended off-axis scattering of pump light
could reduce the contrast ratio between ``on'' and ``off'' states
and thus the performance of the switch. However, this practical
issue might be alleviated using another existing microcavity design
\cite{Diederichs2006} with a lower pump threshold intensity. We
further note that in a fully two-dimensional simulation details of
the switching process (e.g., the characteristic switching time)
might change compared to the results in Fig.~\ref{Figswitch}.
Therefore, simulations beyond the hexagonal geometry considered in
this work are desirable for the future.

\textsc{Conclusions} -- We propose all-optical switching in planar
semiconductor microcavities. Being based on four-wave mixing
instabilities, the proposed scheme is similar to that recently
demonstrated in an atomic vapor \cite{Dawes2005}. Using a
microscopic many-particle theory, our numerical simulations show
that in the studied microcavity system a strong optical beam can be
controlled with a weaker one. Besides its experimental verification,
for the future it is worthwhile to study the influence of vectorial
polarization effects \cite{Schumacher2007c} on the switching
process, and to extend the analysis to other systems
\cite{Diederichs2006,Schumacher2007d}.

\textsc{Acknowledgments} -- This work has been supported by ONR,
DARPA, JSOP. \mbox{S. Schumacher} gratefully acknowledges support by
the Deutsche Forschungsgemeinschaft through project No.
SCHU~1980/3-1. Part of the computation was done at CCIT, University
of Arizona.

%
%

\end{document}